\documentclass[a4paper,11pt]{article}

\pdfoutput=1 
\usepackage{./OmegaDeformationFiles/jheppub_mod}
\usepackage[T1]{fontenc}
\usepackage{lmodern}
\usepackage[utf8]{inputenc}

\usepackage[T1]{fontenc}
\usepackage[utf8]{inputenc}
\usepackage[english]{babel}
\usepackage{graphicx}
\usepackage{amssymb}
\usepackage{mathrsfs}
\usepackage{microtype}
\usepackage{booktabs}
\usepackage{array, tabularx}
\usepackage{chngpage}
\usepackage{braket}
\usepackage{bbm}
\usepackage{mathtools}
\usepackage{amsthm}
\usepackage[english]{varioref}
\usepackage{url}
\usepackage{multicol, multirow}
\usepackage{mparhack}
\usepackage{emptypage}
\usepackage{setspace}
\usepackage{dsfont}
\usepackage{rotating}
\usepackage{enumitem}
\usepackage{comment}
\usepackage{tikz,tikz-cd}
\usepackage{xcolor}

\usepackage[colorlinks=true
,urlcolor=blue
,anchorcolor=blue
,citecolor=blue
,filecolor=blue
,linkcolor=blue
,menucolor=blue
,linktocpage=true
,pdfproducer=medialab
,pdfa=true
]{hyperref}

\usetikzlibrary{cd,shadings}

\numberwithin{equation}{section}  

\newcommand{\beq}{\begin{equation}}
\newcommand{\eeq}{\end{equation}}
\newcommand{\nn}{\nonumber}

\newcommand{\numberset}{\mathbb}

\newcommand{\R}{\numberset{R}}

\newcommand{\mb}[1]{\mathbf{#1}}
\newcommand{\mc}[1]{\mathcal{#1}}

\newcommand{\cA}{\mc{A}}
\newcommand{\cB}{\mc{B}}
\newcommand{\cC}{\mc{C}}
\newcommand{\cD}{\mc{D}}
\newcommand{\cF}{\mc{F}}

\newcommand{\cN}{\mc{N}}

\newcommand{\ph}[1]{\phantom{#1}}
\newcommand{\rd}{{\rm d}}
\newcommand{\dd}{{\rm d}}
\newcommand{\ii}{{\rm i}}
\newcommand{\identity}{\mathds{1}}

\newcommand{\e}{{\rm e}}

\newcommand{\vol}{{\rm vol}}
\newcommand{\hook}{\mathbin{\rule[.2ex]{.4em}{.03em}\rule[.2ex]{.03em}{.9ex}}}

\newcommand{\Ric}{{\rm Ric}}

\allowdisplaybreaks 

\newcommand{\nocontentsline}[3]{}
\newcommand{\tocless}[2]{\bgroup\let\addcontentsline=\nocontentsline#1{#2}\egroup}

\newcommand{\Xb}{X_1}
\newcommand{\Xs}{X_2}

\newcommand{\mexp}{\mathtt{g}}

\newcommand{\los}{\varepsilon}

\newcommand{\Jb}{\mathrm{J}}

\newcommand{\Jcur}{\mathscr{J}}

\newcommand{\aIfour}{a^I_1}
\newcommand{\aIfive}{a^I_2}
\newcommand{\mafour}{\ma_1}
\newcommand{\mafive}{\ma_2}
\newcommand{\Xseven}{X_3}
\newcommand{\Xeight}{X_4}
\newcommand{\mf}{\mathrm{f}}

\newcommand{\JRcur}{\mathbb{J}}
\newcommand{\ma}{\mathrm{a}}

\newlength{\sswidth}
\newcommand{\sla}[1]{
   \settowidth{\sswidth}{$#1$}
   \mbox{$\not{\hspace*{-0.15\sswidth}#1}$}}

\begin{document} 


\begin{center}
\today

\vspace{1cm}

{\Large \bf Topological AdS/CFT and the $\Omega$ deformation}

\vspace{1cm}

{Pietro Benetti Genolini$^a$ and Paul Richmond$^b$}

\vspace{0.8cm}

{$^a${\it Department of Applied Mathematics and Theoretical Physics, \\
University of Cambridge, Wilberforce Road, Cambridge, CB3 OWA, UK}}

\vspace{0.3cm}

{$^b${\it Department of Mathematics, \\
King's College London, Strand, WC2R 2LS, UK}}

\end{center}

\bigskip
\begin{center}
{\bf {\sc Abstract}} 
\end{center}
In this note, we define a holographic dual to four-dimensional superconformal field theories formulated on arbitrary Riemannian manifolds equipped with a Killing vector. Moreover, assuming smoothness of the bulk solution, we study the variation of the holographically renormalized supergravity action in the class of metrics on the boundary four-manifold with a prescribed isometry.

\bigskip


\medskip

\section{Introduction and summary}

Local quantum field theories can be formulated on curved spaces preserving the existence of nilpotent fermionic symmetries by coupling to a background off-shell supergravity and solving the relevant generalized Killing spinor equations in order to have (twisted) covariantly constant supersymmetry parameters \cite{Festuccia:2011ws}. For conformal field theories with supergravity duals, there exists a dual gravity construction: by looking at the conformal boundary of asymptotically locally AdS solutions to a gauged supergravity that is a consistent truncation of ten- or eleven-dimensional supergravity, we realize curved backgrounds of relevant deformations of superconformal field theories \cite{Klare:2012gn}.

Both approaches have been extensively studied in a number of dimensions and with different numbers of conserved supercharges. An extreme case is constructed when the allowed background is an arbitrary manifold, which is achieved via the topological twist \cite{Witten:1988ze}. This case has been famously studied in four dimensions to recover the theory of Donaldson's invariants, and phrased in terms of coupling to four-dimensional $\mc{N}=2$ off-shell supergravity \cite{Karlhede:1988ax, Klare:2013dka}: the associated vector bundle to the $SU(2)_R$ gauge bundle is identified with the bundle of self-dual two-forms on the background manifold, and it is possible to appropriately choose the background $SU(2)_R$ connection in order to cancel the self-dual part of the Levi-Civita spin connection. In this way, effectively a chiral projection of the relevant Killing spinor equation becomes an equation on flat space, and it is always possible to solve it, finding a chiral constant spinor on any curved background. For superconformal field theories, the corresponding construction of the topologically twisted supersymmetric background from gauged five-dimensional supergravity has been considered in \cite{BenettiGenolini:2017zmu}.

A similar construction can be performed on Riemannian manifolds with an isometry: in this case, the Killing vector is used to define the conserved supercharge, and the field theory observables include a sector defining an equivariant cohomology with respect to the isometry. This construction, called the $\Omega$ background, was first introduced on $\R^4$ using rotations in the two transverse planes \cite{Nekrasov:2002qd}, and then presented in generality by Nekrasov and Okounkov on an arbitrary manifold \cite{Nekrasov:2003rj}.\footnote{We will also refer to the general construction as the $\Omega$ deformation.}  Its applications have been numerous and influential, including a crucial r\^ole in the formulation of the AGT correspondence \cite{Alday:2009aq}. Again, as for the topological twist, this construction can be formulated in terms of coupling to a background $\mc{N}=2$ off-shell supergravity \cite{Klare:2013dka}, and one can also formulate the dual supergravity background, as was outlined in the conclusions of \cite{BenettiGenolini:2017zmu}.

In the latter paper, we showed that under certain assumptions of smoothness and existence of the bulk solution, the holographic Ward identity corresponding to the supersymmetric topological twist held. That is, given a curved background with a certain geometric structure on it, $(M_d,g)$, the AdS/CFT dictionary \cite{Witten:1998qj, Gubser:1998bc} associates to it an asymptotically locally hyperbolic solution of an appropriate gauged supergravity theory $(Y_{d+1},G)$ such that $(M_d,[g])$ arises as the conformal boundary. Moreover, in a gravity saddle point approximation
\beq
\label{eq:AdSCFT}
Z[M_d] \ = \ \e^{-S[Y_{d+1}]} \, ,
\eeq
where $Z$ is the partition function of the gauge theory (in a particular limit dependent on the rank of the gauge group), and $S$ is the holographically renormalized supergravity on-shell action. Since for the topologically twisted theory $Z$ is independent of the background Riemannian metric, so should be the on-shell action, as was proved in \cite{BenettiGenolini:2017zmu,BenettiGenolini:2018iuy}. Moreover, the renormalized on-shell action of smooth filling supergravity solutions with the boundary conditions of the topological twist vanishes \cite{BenettiGenolini:2018iuy}.

In this short note, we extend the first computation to an asymptotically locally hyperbolic solution with boundary conditions defining the $\Omega$ background, as we summarised in the conclusions of \cite{BenettiGenolini:2017zmu}. For a four-dimensional $\mc{N}=2$ theory formulated on an $\Omega$ background, we investigate the dependence of the supersymmetric partition function on the Killing vector and on the background metric, showing that it depends on the choice of isometry. 

More specifically, the main difference between the setup here and the setup of \cite{BenettiGenolini:2017zmu} is that there a pair of antisymmetric tensor fields $\mc{B}^{\pm}$ had been consistently set to zero, whereas the boundary conditions of the $\Omega$ background necessarily require them to be non-vanishing. Even though here we study the independence of the on-shell action from variations of the boundary metric preserving an isometry, we do not evaluate the on-shell action for a concrete solution to the theory. It would be very interesting to include the contributions of the $\mc{B}^{\pm}$ fields in the rest of the results of \cite{BenettiGenolini:2018iuy}. For instance, the $\Omega$ background at the boundary requires the existence of two Killing spinors in the bulk. This means that instead of defining a (twisted) $SU(2)$ structure, as it is the case for a single (twisted) spinor in five dimensions, the two of them define an identity structure. Studying this $G$-structure could be instrumental in finding a supersymmetric solution or in evaluating observables for a general class of solutions, assuming their existence. Indeed, in contrast to the Donaldson--Witten twist, an explicit supergravity solution dual to the $\Omega$ deformation on $\mathbb{R}^4$ has been given in \cite{Bobev:2019ylk}. In fact, it is tempting to conjecture that the existence of a Killing vector in the bulk arising from a boundary Killing vector may be a hint into a connection between the equivariant localization at the boundary and the computation of the on-shell action in the supergravity bulk in terms of the contributions from the fixed-point sets of the bulk Killing vector, as in the four-dimensional case \cite{BenettiGenolini:2019jdz}.

Finally, the effective supergravity constructions of \cite{BenettiGenolini:2017zmu,Bobev:2019ylk} and this note are supposed to capture the features of topological subsectors of the physical AdS/CFT duality. It would be interesting to investigate their relation to the twisting constructions of \cite{Costello:2016mgj, Costello:2018zrm}.

\subsection*{Outline}

In section \ref{sec:SUGRA} we very briefly review the structure of the supergravity theory of interest and summarize the relevant steps of the holographic renormalization procedure. In section \ref{sec:ExpansionSUSY} we present the expansion of the supersymmetry equations, and we conclude in section \ref{sec:Variation} by computing the relevant variation of the on-shell action. Throughout the paper, we will heavily rely on the notation and results introduced in our previous work \cite{BenettiGenolini:2017zmu}, to which we refer the reader for some of the details in the computations that we are leaving out of the succinct exposition.

\section{5d \texorpdfstring{$\mathcal{N}=4^+$}{N=4+} supergravity}
\label{sec:SUGRA}

\subsection{Lagrangian and equations of motion}

The five-dimensional gauged supergravity that is relevant to us is the $\mc{N}=4^+$ Romans' theory with gauged $SU(2)\times U(1)$ \cite{Romans:1985ps}. This theory is a consistent truncation of type IIB supergravity on $S^5$ \cite{Lu:1999bw} and of eleven-dimensional supergravity on $N_6$ spaces \cite{Gauntlett:2007sm}. Therefore, the results obtained here apply to $\Omega$ twists of $\mc{N}=4$ SYM and to (some) conformal field theories of class $\mc{S}$ \cite{Gaiotto:2009we} (the choice of theory being dependent on the uplift).\footnote{In fact, a string/M-theoretic construction of the $\Omega$ deformation has been proposed \cite{Hellerman:2011mv, Hellerman:2012zf, Orlando:2013yea}, and it would be interesting to investigate the connections with our work in the truncated five-dimensional gauged supergravity.} In this context, a computation involving a supersymmetric black hole solution to Romans' $\mc{N}=4^+$ theory has already been precisely matched to a supersymmetric Rényi entropy computed in $\mc{N}=4$ SYM \cite{Crossley:2014oea}.

The bosonic dynamical sector of the Euclidean continuation of $\cN=4^+$ Romans' supergravity includes the metric $G_{\mu\nu}$, the dilaton $\phi$, an $SU(2)_R$ gauge field $\cA^I_{\mu}$, a $U(1)_R$ gauge field $\cA_{\mu}$, and two real antisymmetric tensors $B_{\alpha}$ charged under $U(1)_R$.\footnote{More precisely, in order to have real solutions to the equations of motion we need to require $\cA$ to be purely imaginary, so the gauged subgroup is $SO(1,1)$ with connection $\cC=\ii \cA$.}  We adopt the same conventions used in \cite{BenettiGenolini:2017zmu}, and we consistently define the complex combinations $\cB^{\pm} \equiv B^1 \pm \ii B^2$ and the scalar $X\equiv \e^{-\frac{1}{\sqrt{6}}\phi}$. The curvatures are $\cF=\rd \cA$ and $\cF^I = \rd \cA^I - \tfrac{1}{2}\epsilon^{IJK}\cA^J\wedge \cA^K$, and we define the covariant derivative $H^{\pm} = \rd \cB^{\pm} \mp \ii \cA \wedge \cB^{\pm}$.

The Wick-rotated action is
\beq
\label{IEuclid}
\begin{split}
	I \ =  \  - \frac{1}{2\kappa_5^2} \int \ \Big[& R \, {*1} - 3 X^{-2} \dd X \wedge *\dd X + 4  ( X^2 + 2 X^{-1} ) \, {*1} - \tfrac{1}{2} X^4 \, \cF \wedge *\cF  \\
	&\ - \tfrac{1}{4} X^{-2} \, ( \cF^I \wedge * \cF^I + {\mathcal{B}^-} \wedge * \mathcal{B}^+ )+ \tfrac{1}{8} {\mathcal{B}^-} \wedge H^+ - \tfrac{1}{8} \mathcal{B}^+ \wedge {H}^- \\
	&\ - \tfrac{\ii}{4} \cF^I \wedge \cF^I \wedge \cA \Big] \, .
\end{split}
\eeq
The associated equations of motion are
\begin{align}
	\label{Xeom}
	\begin{split}
	\dd ( X^{-1}\, {*\, \dd X})   \  = & \ \tfrac{1}{3} X^4 \, \cF \wedge *\cF - \tfrac{1}{12} X^{-2} \, (\cF^I \wedge *\cF^I + {\mathcal{B}^-} \wedge *\mathcal{B}^+) \\  &  - \tfrac{4}{3}  (X^2 - X^{-1})\, *1~,  \end{split}\\[5pt]
	\dd ( X^{-2} * \cF^I ) \ =& \  \   \epsilon^{I}_{\ JK} X^{-2} * \cF^J \wedge \cA^K - \ii \cF^I \wedge \cF~, \label{AIeom} \\[5pt]
	\dd ( X^4 * \cF ) \ =& \ - \tfrac{\ii}{4} \cF^I \wedge \cF^I - \tfrac{\ii}{4} {\mathcal{B}^-} \wedge \mathcal{B}^+~, \label{Aeom} \\[5pt]	
	H^\pm \ =& \ \pm X^{-2} * \mathcal{B}^\pm \label{Beom}~, \\[5pt]
	%
	%
	%
	%
	\label{geom}
	\begin{split}
		R_{\mu\nu}  \ = & \  3  X^{-2} \partial_\mu X \partial_\nu X - \tfrac{4}{3} ( X^2 + 2 X^{-1} ) G_{\mu\nu}+ \tfrac{1}{2} X^4 \big( \cF_\mu{}^\rho \cF_{\nu\rho} - \tfrac{1}{6} G_{\mu\nu} \cF^2 \big) \\
	& + \tfrac{1}{4} X^{-2} \big( \cF^I_\mu{}^\rho \cF^I_{\nu\rho} - \tfrac{1}{6} G_{\mu\nu} ( \cF^I )^2 + {\mathcal{B}^-}_{(\mu}{}^\rho \mathcal{B}^+_{\nu)\rho} - \tfrac{1}{6} G_{\mu\nu} {\mathcal{B}^-}_{\rho\sigma} \mathcal{B}^{+ \rho\sigma} \big) \, ,
	\end{split}
\end{align}
where $\cF^2\equiv \cF_{\mu\nu}\cF^{\mu\nu}$, $( \cF^I )^2\equiv \sum_{I=1}^3\cF^I_{\mu\nu}\cF^{I\mu\nu}$.

The four gravitini and four dilatini present in the Lorentzian theory transform in the $\mb{4}$ of the global R-symmetry group $Sp(2)\cong Spin(5)$, and so does the spinor supersymmetry parameter $\epsilon$. However, since we have gauged the subgroup $SU(2)_R\times U(1)_R$, we naturally split the generators of the Clifford algebra Cliff$(5,0)$ corresponding to $Spin(5)$ into $\Gamma^I$, $I=1,2,3$, on which $SU(2)$ acts in the $\mb{3}$, and $\Gamma^{\alpha}$, $\alpha=4,5$, on which $U(1)$ acts in the $\mb{2}$. The condition to have a supersymmetric solution is the vanishing of the variations of the gravitini and dilatini, which in Euclidean signature read
\begin{align}
	\label{BulkGravitino}
	\begin{split}
0\ =& \ D_\mu \epsilon + \tfrac{\ii}{3} \gamma_\mu \Big( X + \tfrac{1}{2} X^{-2} \Big)  \Gamma_{45}  \epsilon \\
	& \ + \tfrac{\ii}{24} ( \gamma_\mu{}^{\nu\rho} - 4 \delta_\mu^\nu \gamma^\rho ) \left( X^{-1} \big(\cF_{\nu\rho}^I\Gamma_I + B^\alpha_{\nu\rho} \Gamma_\alpha  \big) + X^2 \mathcal{F}_{\nu\rho} \right) \epsilon~, 
	\end{split} \\
	\label{BulkDilatino}
	\begin{split}
0\  =& \ \tfrac{\sqrt{3}}{2} \ii \gamma^\mu  X^{-1} \partial_\mu X \epsilon + \tfrac{1}{\sqrt{3}} \Big( X - X^{-2} \Big)  \Gamma_{45}  \epsilon  \\
	& \ + \tfrac{1}{8\sqrt{3}} \gamma^{\mu\nu} \Big( X^{-1} \big( \cF_{\mu\nu}^I  \Gamma_I   + B^\alpha_{\mu\nu} \Gamma_\alpha  \big) - {2} X^2 \mathcal{F}_{\mu\nu} \Big) \epsilon  \, .
	\end{split}
\end{align}
Here the gauge covariant derivative is
\begin{align}
D_{\mu}\epsilon \ \equiv \ \nabla_{\mu}\epsilon + \tfrac{1}{2} \cA_{\mu}\Gamma_{45}\epsilon + \tfrac{1}{{2}} \cA^I_{\mu}\Gamma_{I45}\epsilon~,
\end{align}
and $\gamma_{\mu}$ generate the spacetime Clifford algebra. For the R-symmetry Clifford algebra, we choose the following generators
\begin{align}
\Gamma^I \ = \ \sigma_3\otimes \sigma_I \, , \qquad \Gamma_4 \ = \ \sigma_1\otimes 1_2~, \qquad \Gamma_5 \ = \ 
\sigma_2 \otimes 1_2 \, ,
\end{align}
where $\sigma_I$ are the Pauli matrices, so that we may write
\beq
\epsilon \ = \ \begin{pmatrix} \epsilon^+ \\ \epsilon^- \end{pmatrix}~,\label{pmeigen}
\eeq
denoting by $\epsilon^{\pm}$ the projection onto the $\pm \ii$ eigenspaces 
of $\Gamma_{45}$, respectively. In this way, there is a natural splitting of the equations between the two eigenspaces.

\subsection{Perturbative expansion}

For an asymptotically locally hyperbolic solution, we may assume that the fields have a Fefferman--Graham expansion in a neighbourhood of the conformal boundary in terms of a radial coordinate $z$.

We take the metric to have the form
\begin{align}
G_{\mu\nu}\rd x^\mu \rd x^\nu \ = \ \frac{1}{z^2} \dd z^2 + \frac{1}{z^2} \mexp_{ij} \dd x^i\dd x^j \ = \ \frac{1}{z^2} \dd z^2 + h_{ij} \dd x^i\dd x^j \, ,
\end{align}
and we assume the expansions
\begin{align}
\label{metricexp} 
\mexp_{ij} \ =& \ \mexp_{ij}^{0}+z^2 \mexp_{ij}^{2}+z^4 \big( \mexp_{ij}^{4} + h_{ij}^{0} ( \log z )^2 + h_{ij}^{1} \log z \big) + o(z^4)~, \\[5pt]
\label{eq:ExpansionB}
\begin{split}
\mc{B}^{\pm} \ =& \ \frac{1}{z}b^\pm + \dd z \wedge B^\pm_1 + z(b_2^\pm \log z + b^\pm_3)  +  z \, \rd z \wedge (B_2^{\pm} \log z + B_3^{\pm}) \\
& \ +z^2 (b_4^{\pm}\log^2 z + b_5^{\pm}\log z+ b_6^{\pm}) + z^2\rd z \wedge (B_4^{\pm}\log^2 z + B_5^{\pm} \log z + B_6^{\pm}) \\
& \ + o(z^2) \, ,
\end{split}\\[5pt]
\label{eq:ExpansionA}
\mc{A} \ =& \ \ma + z^2 (\ma_1 \log z + \ma_2) + o(z^3) \, , \\[5pt]
X \ =& \ 1 + z^2 \left( \Xb \log z + \Xs \right) + z^4 (\Xseven \log z + \Xeight ) + o(z^4)~, \label{Xexp} \\[5pt]
	\cA^I \ =& \ A^I + z^2 (\aIfour \log z + \aIfive)  + o(z^2)~. \label{aIexp}
\end{align}
Note that we have already used gauge freedom to remove some of the fields that would be present in the most generic expansion, and set to zero those that vanish because of the equations of motion.
We would then substitute these forms into the equations of motion \eqref{Xeom}-\eqref{geom} and find relations between the coefficients of the expansion. This is done in generality in \cite{BenettiGenolini:2017zmu}. Here, we will summarize the results with the boundary conditions fixed to be the twisted ones.

In fact, we assume that the spinor has a Fefferman--Graham-like expansion in a neighbourhood of the conformal boundary
\begin{equation}
\label{eq:SpinorFG}
	\begin{split}
	\epsilon^\pm \ =& \ z^{-1/2}\los^\pm + z^{1/2}\eta^\pm + z^{3/2} ( \log z \, \tilde{\varepsilon}^{3,\pm} + \varepsilon^{3,\pm} ) + z^{5/2} ( \log z \, \tilde{\varepsilon}^{5,\pm} + \varepsilon^{5,\pm} ) \\
	&+ z^{7/2} \big( (\log z)^2 \, \mathring{\varepsilon}^{7,\pm} + \log z \, \tilde{\varepsilon}^{7,\pm} + \varepsilon^{7,\pm} \big) + o(z^{7/2}) \, ,
	\end{split}
\end{equation}
and substitute this form in the generalized Killing spinor equations \eqref{BulkGravitino} and \eqref{BulkDilatino}. At the lowest order in $z$ we obtain the boundary generalized Killing spinor equations
\begin{equation}
	\label{eq:BoundaryGravitino}
	 \mathcal{D}^{(0)}_i\los^\pm - \tfrac{\ii}{4}b^\pm_{{i}{j}}\gamma^{j}\los^\mp \mp \gamma_{i}\eta^\pm  \ = \ 0~, 
\end{equation}
where the covariant derivative is
\begin{equation}
\mathcal{D}^{(0)}_i \ \equiv \  \nabla^{(0)}_{{i}} \pm \tfrac{\ii}{2}\ma_{{i}} + \tfrac{\ii}{{2}} A_{{i}}^I\sigma_I~,
\end{equation}
and the boundary dilatino equations
\begin{equation}
\label{eq:BoundaryDilatino}
\sla{\mathcal{D}^{(0)}}\sla{\mathcal{D}^{(0)}}\los^\pm - \ii \mathcal{D}_{{i}}(b^\pm)^{{i}}_{\ {j}}\gamma^{{j}}\los^\mp 
+ \left(4\Xb + \tfrac{1}{3}R\right)\los^\pm \mp 2\ii\,  \mf \cdot \los^\pm \ =\  0~.
\end{equation}
These equations correspond to the supersymmetry equations for off-shell Euclidean $\cN=2$ conformal supergravity in four dimensions, which have been originally  studied in the context of rigid supersymmetric backgrounds in \cite{Gupta:2012cy, Klare:2013dka}.

As already discussed in \cite{BenettiGenolini:2017zmu} in general, and in \cite{Klare:2013dka, Bobev:2019ylk} for the specific case of the $\Omega$ background on $\R^4$, we may solve the equations \eqref{eq:BoundaryGravitino} and \eqref{eq:BoundaryDilatino} on a Riemannian four-manifold $(M_4,g,\xi)$ with an isometry generated by the Killing vector $\xi$ by setting  
\begin{equation}
\label{eq:NOTwist}
\begin{split}
\Xb \ = \ - \tfrac{1}{12}R \, , \quad \ma \ = \ 0 \, , \quad b^- \ = \ 0 \, , \quad b^+ \ = \ 2 (\rd \xi^{\flat})^- \, , \\
\cD^{(0)}_i\los^+ \ = \ 0 \, , \quad \los ^- \ = \ \ii \xi^\flat \cdot \los^+ \, , \quad \eta^+ \ = \ 0 \, , \quad \eta ^- \ = \ - \tfrac{\ii}{4}\rd\xi^\flat \cdot \los^+ \, ,
\end{split}
\end{equation}
where $R=R(g)$ is the curvature scalar of the boundary metric $g \equiv \mexp^0$, $\flat$ is the musical isomorphism using $g$ and $(\rd\xi^{\flat})^-$ is the anti-self-dual part of $\rd\xi^\flat$ with respect to the Hodge dual defined by the boundary metric.
We also introduce the Clifford product of a $k$-form $\omega$ and a spinor $\epsilon$ as $\omega\cdot \epsilon \equiv \frac{1}{k!}\omega_{i_1\dots i_k}\gamma^{i_1\dots i_k}\epsilon$. Notice that, differently from the case of the topological twist, $b^+\neq 0$ and so are $\los^- , \eta^-$. However, finding a covariantly constant spinor $\los^+$ again requires identifying the $SU(2)_R$ gauge bundle with the self-dual part of the spin connection. This allows $\epsilon$ to exist on an arbitrary Riemannian manifold, even if not spin, as it becomes a section of the tensor product bundle $\mc{S}^+\otimes \mc{V}$, where $\mc{S}^+$ is, on a spin manifold, the positive chirality spin bundle, and $\mc{V}$ is the rank 2 vector bundle associated to the $SU(2)_R$ gauge bundle. Thus, we have a $Spin_{SU(2)}$ structure.\footnote{These $G$-structures originally appeared in the supergravity literature \cite{Hawking:1977ab, Back:1978zf, Avis:1979de}, and have more recently been used also in the context of phases of quantum field theories, e.g. \cite{Cordova:2018acb, Wang:2018qoy}.} Concretely, choosing the $\gamma$ matrices
\begin{align}
\gamma_{\bar{a}} \ = \  \left(\begin{matrix} 0 & \ii \sigma_{\bar{a}} \\ -\ii \sigma_{\bar{a}} & 0\end{matrix}\right)~, \quad 
\gamma_{\bar{4}} \ = \ \left(\begin{matrix} 0 & -\identity_2 \\ -\identity_2 & 0\end{matrix}\right)~, \quad 
\gamma_{\bar{z}} \ =\   \left(\begin{matrix} \identity_2 & 0 \\ 0 & -\identity_2\end{matrix}\right)~,\label{gammamatrices}
\end{align}
we find that $\mc{D}^{(0)}_i\los^+=0$ is solved by the following choice
\beq
\label{eq:TopTwist}
A^I_i = \tfrac{1}{2}\Jb^I_{jk}(\omega_i^{(0)})^{jk} \, , \qquad (\los^+)^i_{\ph{i}\alpha} = (\ii\sigma_2)^i_{\ph{i}\alpha}c \, ,
\eeq
where $\omega^{(0)}$ is the spin connection on the boundary, and $\Jb^I_{ij} = \eta^I_{\overline{ij}}\e^{\overline{i}}_i \e^{\overline{j}}_j$ is a triplet of globally $SU(2)$-twisted self-dual two-forms which in a vierbein basis have the same components as the self-dual 't Hooft symbols. Moreover, we have indices $i=1,2$ for the doublet of spinors and $\alpha =1, 2$ for the positive chirality components, and we choose $c\in\R$ to have a symplectic Majorana spinor.

From now on, to simplify notation in the remainder of this note, we will drop the superscript $(0)$ as all geometrical quantities will be with respect to the boundary metric. With the choice of $SU(2)$ gauge field \eqref{eq:TopTwist}, the self-dual two-forms $\Jb^I$ satisfy a number of identities that will be relevant for our later computations, including
\begin{align}
\label{outerJI}
\Jb^I_{ij}\Jb^I_{kl} \ =& \ g_{ik}g_{jl} - g_{il}g_{jk} + \epsilon_{ijkl}~, \\
\label{QKeqn}
\nabla_i \Jb^I_{jk} \ =& \  \epsilon^I_{\ JK} A^J_i \Jb^K_{jk}~.
\end{align}

Moreover, identifying the connection on the (vector bundle associated to the) gauge bundle and the self-dual part of the spin connection implies a relation between the two curvatures, which takes the form
\beq
\label{Ftwist}
F^I_{ij} \ = \ \tfrac{1}{2}\Jb^I_{kl} R_{ij}^{\ \ kl} \, ,
\eeq
where $R_{ijkl}$ is the boundary Riemann tensor.

With the boundary conditions \eqref{eq:NOTwist} that define a supersymmetric background compatible with the generic Nekrasov--Okounkov twist, the antisymmetric tensors have the following expansions
\begin{align}
\begin{split}
\mc{B}^{+} \ =& \ \frac{1}{z}2(\rd\xi^\flat)^- + \dd z \wedge (-2\xi \hook \Ric) + z(b_2^+ \log z + b^+_3)   \\
& \ +z^2 (b_5^{+}\log z+ b_6^{+}) + z^2\rd z\wedge ( B_5^{+} \log z + B_6^+) + o(z^2) \, ,
\end{split} \\[5pt]
\begin{split}
\mc{B}^{-} \ =& \ z\, b^-_3  +  z^2\dd z\wedge *\rd b_3^- + o(z^2) \, ,
\end{split}
\end{align} 
where the terms on the right hand side satisfy a number of equations that can be determined by expanding \eqref{Beom}, including
\begin{align}
(b_{2}^{+})^- \ =& \ -\tfrac{1}{6} R\, (\rd\xi^\flat)^- \, , \\[5pt]
(b_3^+)^- \ =& \ - \frac{1}{2}b_2^+ + \frac{8X_2 + R}{4}(\rd\xi^\flat)^- + * \left( \Ric \circ (\rd\xi^\flat)^- \right) - \rd \left( \xi \hook \Ric \right) \\[5pt]
B_5^+ \ =& \ - *\dd b_2^+ - \tfrac{1}{3}R\, \xi \hook \Ric \, , \\[5pt]
*b^-_3 \ =& \ - b_3^- \, ,
\end{align} 
here $(\Ric\circ (\rd\xi^\flat)^-)_{ij} \equiv \Ric_{[i}^{\ph{[i}k}(\rd\xi^\flat)^-_{|k|j]}$, and we have not written the next few equations at subleading orders as they are not relevant for our purposes.

\subsection{Holographic renormalization}

The holographic renormalization of the divergences of the on-shell action has been considered in full generality\footnote{In this subsection we do not apply the boundary conditions \eqref{eq:NOTwist}.} in \cite{BenettiGenolini:2017zmu} (with earlier work on the Lorentzian version of the theory in \cite{Ohl:2010au}). We evaluate the Euclidean action \eqref{IEuclid} on a solution, add the Gibbons--Hawking--York term and counterterms required to cancel the divergences \cite{Emparan:1999pm, deHaro:2000vlm, Taylor:2000xw}, and find the value of the on-shell action in the limit where we remove the cutoff $\delta$
\beq
\label{eq:RenAction}
S \ = \ \lim_{\delta\to 0}\left( I_{\rm on-shell} + I_{\rm GHY} + I_{\rm counterterm} \right) \, .
\eeq
Having the finite on-shell action, we can compute the boundary VEVs
\beq
\label{VEVs}
\begin{split}
	\langle T_{ij} \rangle \ = \ \frac{2}{\sqrt{g}} \frac{\delta S}{ \delta g^{ij} } \, , \, \, \qquad &\qquad \langle \Xi \rangle \ = \ \frac{1}{\sqrt{g}} \frac{\delta S}{ \delta \Xb } \, , \\  
	\langle \Jcur_I^{i} \rangle \ =  \ \frac{1}{\sqrt{g}} \frac{\delta S}{ \delta A^I_i }~, 
	\qquad \langle \, \JRcur^i \rangle \ =& \ \frac{1}{\sqrt{g}} \frac{\delta S}{ \delta \ma_i }~, \qquad \langle \Upsilon^{\pm,ij} \rangle \ = \ \frac{2}{\sqrt{g}}\frac{\delta S}{\delta b^{\pm}_{ij}} \, .
\end{split}
\eeq
A straightforward computation then leads to the following finite expressions
\begin{align}
	\langle T_{ij} \rangle \ =& \   \frac{1}{\kappa_5^2} \bigg[ 2 \mexp^4_{ij} + \tfrac{1}{2} h^1_{ij} - \tfrac{1}{2} g_{ij} ( 4 t^{(4)} - 2 t^{(2,2)} + u^{(1)} )  - 3 g_{ij} \Xs^2  - 3 g_{ij} X_1X_2 - \mexp_{ij}^{2} t^{(2)}\nn \\
	&\qquad + \tfrac{1}{4} \Big( \nabla^k \nabla_i \mexp^2_{jk} + \nabla^k \nabla_j \mexp^2_{ik} - \nabla^2 \mexp^2_{ij} - \nabla_i \nabla_j t^{(2)} \Big) - \tfrac{1}{4} g_{ij} \left( \nabla^k \nabla^l \mexp^2_{lk} - \nabla^2t^{(2)} \right) \nn \\
	&\qquad + \tfrac{1}{4} g_{ij} \big( \mexp^2_{kl} \mathscr{R}^{kl} \big) - \tfrac{1}{4} \mexp_{ij}^{2} {R}
	 \nn \\
	&\qquad - \tfrac{1}{8} \big[ (b^+)_{(i}{}^k (b^-_3)_{j)k} + (b^-)_{(i}{}^k (b^+_3)_{j)k} - \tfrac{1}{4} g_{ij} \big(  \langle {b}^+ , {b}_3^- \rangle + \langle {b}^- , {b}_3^+ \rangle \big) \big] \nn \\
	&\qquad + \tfrac{1}{8} (b^+)_{(i|k|} (\mexp^2)^{kl} ({b}^-)_{j)l} \bigg] \label{Tij} \, , \\
	\langle \Xi \rangle \ =& \  \frac{3}{\kappa_5^2} \Xs \, , \label{Xi}\\
	\langle \Jcur^{I}_i \rangle \ =&  \ -\frac{1}{4\kappa_5^2}\left[(\aIfour)_i+2(\aIfive)_i - \ii \big( *(\ma\wedge F^I) \big)_i \right]\label{calJi} \, ,\\
	\langle \, \JRcur_i \rangle \ =&  \ -\frac{1}{2\kappa_5^2}\left[(\mafour)_i+2(\mafive)_i \right]\label{bbJ} \, , \\
	\langle \Upsilon^{\pm,ij} \rangle \ =& \ \frac{1}{16\kappa^2_5}\left[ \tfrac{1}{2}t^{(2)}(b^{\mp})^{ij} \mp 2(*(\mexp^2\circ b^{\mp}))^{ij} - (b_3^{\mp})^{ij} \pm (* b_3^{\mp})^{ij} \right] \, , \label{Ypm} 
\end{align}
where we have defined $\langle \alpha , \beta  \rangle = \alpha_{i_1 \cdots i_p} \beta^{i_1 \cdots i_p}$ and $\mathscr{R}_{ij} = R_{ij} - \frac{1}{4} (b^+)_{(i}{}^k (b^-)_{j)k}$.
As standard in AdS/CFT, these expressions contain a number of terms that are not determined by the boundary conditions and the perturbative expansion of the equations of motion: $\mexp^4_{ij}, X_2, a^I_2, \ma_2, b_3^{\pm}$.

Having the expressions for the one-point functions, we may consider specific variations of the fields and compute the holographic Ward identities \cite{Witten:1998qj, Henningson:1998gx}. For instance, we find that the Weyl anomaly takes the form
\beq
\label{eq:WeylAnomaly}
\begin{split}
\mc{A}_W \ &= \ - \frac{1}{\kappa^2_5}\left( \langle T^i_{\ph{i}i} \rangle + 2 \langle \Xi \rangle X_1 - \frac{1}{2}\langle \Upsilon^+,b^+ \rangle - \frac{1}{2}\langle \Upsilon^-,b^- \rangle \right) \\
&= \ -\frac{1}{8\kappa_5^2}\left[ \mathscr{R}_{ij}\mathscr{R}^{ij} - \frac{1}{3}\mathscr{R}^2 + 12 X_1^2 - \langle \mf,\mf \rangle - \frac{1}{2}\langle F^I, F^I \rangle - \langle \mc{D}b^+,\mc{D}b^- \rangle \right] \, .
\end{split}
\eeq
Notice that the gravitational part reproduces the standard expression for the Weyl anomaly of a four-dimensional superconformal field theory -- corrected by the $b^{\pm}$ fields -- since $R_{ij}R^{ij} - \frac{1}{3}R^2 = \frac{1}{2}(C^2-\mc{E})$. Moreover, the expression corresponds to that obtained in \cite{Ohl:2010au} for the Lorentzian version of the theory and to the bosonic part of the Lagrangian of $\mc{N}=2$ conformal supergravity \cite{Bergshoeff:1980is}.

Analogously, one can compute the holographic Ward identity corresponding to the boundary R-symmetry by performing a gauge transformation at the boundary
\beq
- \ii * \rd * \langle \JRcur \rangle \ = \ \frac{1}{2} \big\langle \langle \Upsilon^+ \rangle, b^+ \big\rangle - \frac{1}{2} \big\langle \langle \Upsilon^- \rangle, b^- \big\rangle \, .
\eeq
This can be equivalently expressed using the one-point function \eqref{bbJ} and the form of $\ma_1, \ma_2$ fixed by the equations of motion as a constraint
\beq
\rd \left( b^+ \wedge *\mc{D}b^- - b^- \wedge *\mc{D}b^+ \right) + 2  F^I \wedge F^I \ = \ 0  \, .
\eeq
With the boundary conditions of the Nekrasov--Okounkov twist \eqref{eq:NOTwist}, we find the same integrated condition as in \cite{BenettiGenolini:2017zmu} in terms of the Euler characteristic $\chi$ and signature $\sigma$ of the boundary four-manifold
\beq
\label{eq:TopConstraint}
2\chi(M_4) + 3\sigma(M_4) \ = \ 0 \, .
\eeq
This becomes a constraint on the topology of the boundary in order to have a smooth filling. Notice that in both the topological and the Nekrasov--Okounkov twists, the expression for Weyl anomaly \eqref{eq:WeylAnomaly} reduces to
\beq
\mc{A}_W \ = \ \frac{1}{32\kappa_5^2}\left(\mc{E}+\mc{P}\right) \, ,
\eeq
which when integrated imposes the same constraint as the vanishing of the $U(1)_R$ anomaly equation, \eqref{eq:TopConstraint}.

Finally, for completeness, one can compute the divergence of the holographic $SU(2)_R$-current, for which we find
\beq
\mc{D}*\langle \mathscr{J}^I \rangle \ = \ -\frac{\ii}{4\kappa^2_5} \mf \wedge F^I \, .
\eeq

\section{Expansion of the supersymmetry equations}
\label{sec:ExpansionSUSY}

In order to evaluate the holographic Ward identity corresponding to the variation of the action, we need additional relations between the subleading terms in the expansion of the fields \eqref{metricexp}-\eqref{aIexp}, which we find from the expansion in the bulk of the supersymmetry equations \eqref{BulkGravitino} and \eqref{BulkDilatino}. It is particularly useful to project the two equations on the two eigenspaces of $\Gamma_{45}$, using the splitting \eqref{pmeigen}. We find that the projections of the dilatino equations \eqref{BulkDilatino} are
\begin{align}
\label{eq:DilatinoPlus}
	\begin{split}
	0 \ =& \ \tfrac{\sqrt{3}}{2} \ii \gamma^\mu X^{-1} \partial_\mu X \epsilon^+ + \tfrac{\ii}{\sqrt{3}} \big( X - X^{-2} \big) \epsilon^+ \\
	&+ \tfrac{1}{8\sqrt{3}} \gamma^{\mu\nu} \big( X^{-1} \cF_{\mu\nu}^I \sigma_I \epsilon^+ + X^{-1} \mathcal{B}^+_{\mu\nu} \epsilon^- - {2} X^2 \mathcal{F}_{\mu\nu} \epsilon^+ \big) \, ,
	\end{split} \\[5pt]
\label{eq:DilatinoMinus}
\begin{split}
	0 \ =& \ \tfrac{\sqrt{3}}{2} \ii \gamma^\mu X^{-1} \partial_\mu X \epsilon^- - \tfrac{\ii}{\sqrt{3}} \big( X - X^{-2} \big) \epsilon^- \\
	&+ \tfrac{1}{8\sqrt{3}} \gamma^{\mu\nu} \big( - X^{-1} \cF_{\mu\nu}^I \sigma_I \epsilon^- + X^{-1} \mathcal{B}^-_{\mu\nu} \epsilon^+ - {2} X^2 \mathcal{F}_{\mu\nu} \epsilon^- \big)  \, , \end{split}
\end{align}
whilst the projection of the gravitino equations \eqref{BulkGravitino} are
\begin{align}
\label{eq:BulkGravitinoPlus}
	\begin{split}
	0 \ =& \ \partial_{\mu} \epsilon^+ + \tfrac{1}{4} \Omega_\mu{}^{mn} \gamma_{mn} \epsilon^+ + \tfrac{\ii}{2} \cA_{\mu}\epsilon^+ + \tfrac{\ii}{{2}} \cA^I_{\mu}\sigma_{I}\epsilon^+ - \tfrac{1}{3} \gamma_\mu \big( X + \tfrac{1}{2} X^{-2} \big) \epsilon^+ \\
	& \ + \tfrac{\ii}{24} ( \gamma_\mu{}^{\nu\rho} - 4 \delta_\mu^\nu \gamma^\rho ) \big( X^{-1} \cF_{\nu\rho}^I \sigma_I \epsilon^+ + X^{-1} \mathcal{B}^+_{\nu\rho} \epsilon^- + X^2 \mathcal{F}_{\nu\rho} \epsilon^+ \big)
	\end{split} \\[10pt]
	\begin{split}
	0 \ =& \ \partial_{\mu} \epsilon^- + \tfrac{1}{4} \Omega_\mu{}^{mn} \gamma_{mn} \epsilon^- - \tfrac{\ii}{2} \cA_{\mu}\epsilon^- + \tfrac{\ii}{{2}} \cA^I_{\mu}\sigma_{I}\epsilon^- + \tfrac{1}{3} \gamma_\mu \big( X + \tfrac{1}{2} X^{-2} \big) \epsilon^- \\
	& \ + \tfrac{\ii}{24} ( \gamma_\mu{}^{\nu\rho} - 4 \delta_\mu^\nu \gamma^\rho ) \big( - X^{-1} \cF_{\nu\rho}^I \sigma_I \epsilon^- + X^{-1} \mathcal{B}^-_{\nu\rho} \epsilon^+ + X^2 \mathcal{F}_{\nu\rho} \epsilon^- \big)  \, .
\end{split}
\end{align}

Notice that the $\cB^{\pm}$ terms mix the two eigenspaces: in \cite{BenettiGenolini:2017zmu}, $\cB^{\pm}\equiv 0$ and we could work with the consistently truncated theory where $\epsilon^- \equiv 0$. This is not true in this case, as already clear from the leading order terms in the expansion of the spinor in \eqref{eq:NOTwist}. However, the components of the spinor in the $-\ii$ eigenspace of $\Gamma_{45}$ are determined, at least to the order relevant for us, by those in the $+\ii$ eigenspace (for instance, we see this already in \eqref{eq:NOTwist}). Order by order, it is possible to prove that this statement is true and that the $\epsilon^-$ component of the equation reduces to some geometric identity involving the Killing vector.

The expansion of the equations is analogous to what had been studied in \cite{BenettiGenolini:2017zmu}, so here we will be very brief, referring the reader to the previous paper for more detail. From the first few orders of the bulk dilatino on $\epsilon^+$ \eqref{eq:DilatinoPlus}, we find the following relations between the bosonic fields
\begin{align}
a^I_1\hook\Jb^I \ =& \ \frac{1}{4}\rd R \, , \\
a^I_2\hook \Jb^I \ =& \ -2\ii\ma_2 - \frac{1}{8}\rd R - 3 \, \rd X_2 + \xi \hook b_3^- \, , \\
\label{eq:covDX2}
3\nabla_i\nabla_jX_2 \ =& \ \mc{D}_{(i}(a_2^I)^k\Jb^I_{j)k} - 2\ii \nabla_{(i}(\ma_2)_{j)} - \frac{1}{8}\nabla_i\nabla_j R + \nabla_{(i}(\xi \hook b_3^-)_{j)} \, .
\end{align}
The contraction of the latter leads to
\beq
\label{eq:LaplacianX2}
3\nabla^2 X_2 \ = \  \frac{1}{2}\langle \mc{D}a^I_2,\Jb^I \rangle - \frac{1}{16}(\mc{E}+\mc{P}) - \frac{1}{4}\langle (\rd\xi^\flat)^-, b_3^- \rangle - \frac{1}{8}\nabla^2 R +\langle \xi^\flat, *\rd b_3^- \rangle \, .
\eeq

We then expand the bulk gravitino on $\epsilon^+$ \eqref{eq:BulkGravitinoPlus}, both in the radial direction and along the boundary, with the $SO(4)$ gauge choice of frame $(\e^{(2)})^{\overline{i}}_i = \frac{1}{2}(\mexp^2)^{\overline{i}}_{\ph{i}\overline{j}} \e^{\overline{j}}_i$, $(\e^{(2)})_{\overline{i}}^i = -\frac{1}{2}\e_{\overline{j}}^i (\mexp^2)^{\overline{j}}_{\ph{i}\overline{i}}$. We then find a few expressions for the fields that are already fixed by the expansions of the bosonic equations of motion, and 
\begin{align}
\label{eq:h1SUSY}
\begin{split}
h^1_{ij} \ &= \ \tfrac{1}{192}g_{ij}R^2 + \tfrac{1}{12}g_{ij} RX_2 - \tfrac{1}{24}\nabla_i\nabla_jR - \tfrac{1}{48}g_{ij}\nabla^2 R \\
& \ \ \ \ \ - \tfrac{1}{8}\left( R_i^{\ph{i}k}R_{jk} + R_{iklj}R^{kl} - \nabla^2 R_{ij} - \tfrac{1}{2}\epsilon_{kmn(j}R^{kl}R^{mn}_{\ph{mn}i)l} \right) \, , 
\end{split} \\[10pt]
\label{eq:g4h1SUSY}
\begin{split}
4\mexp^4_{ij} + h^1_{ij} \ &= \ 2\nabla_i\nabla_j \left( X_2 + \tfrac{1}{24}R \right) + 2\ii \nabla_{(i}(\ma_2)_{j)} + \left(X_2 - \tfrac{1}{12}R \right)R_{ij} \\
& \ \ \ \ + g_{ij} \left( - \tfrac{1}{6}RX_2 - 2X_2^2 + \tfrac{1}{12}R_{ij}R^{ij} \right) + \tfrac{1}{4}R_{ik}R^k_{\ph{k}j} \\
& \ \ \ \ - \tfrac{1}{8}\epsilon^{mnk}_{\ph{mnk}j}R_{mnli}R_k^{\ph{k}l} + \tfrac{1}{4}R_{iklj}R^{kl} + \tfrac{1}{3}\left[2\mc{D}a^I_2 - *\mc{D}a^I_2\right]_{(i|k|}\Jb^{Ik}_{\ph{Ik}j)} \\
& \ \ \ \ + \tfrac{1}{2}(b_3^-)_{(i}^{\ph{(i}k}(\rd\xi^\flat)^{+}_{j)k} + \tfrac{1}{3}\langle *\rd b_3^-, \xi^\flat \rangle \, g_{ij} - \xi^\flat_{(i}(*\rd b_3^-)_{j)} - \nabla_{(i}\left(\xi\hook b^-_3\right)_{j)} \, .
\end{split}
\end{align}
Showing the equivalence of the expansion of the bosonic and fermionic equations of motion requires a number of identities from differential geometry, as already pointed out in \cite{BenettiGenolini:2017zmu}, and a few manipulations of the differential forms based on their duality properties. Here we mention one that is particularly useful for the following as well: for two anti-self-dual two-forms $\alpha, \beta$
\beq
\label{eq:UsefulIdentity}
\alpha_{(i}^{\ph{(i}k}\beta_{j)k} = \frac{1}{4}\langle \alpha, \beta \rangle \, g_{ij} \, .
\eeq

\section{Variation of the action}
\label{sec:Variation}

The holographic Ward identity for the variation of the renormalised on-shell action \eqref{eq:RenAction} with respect to a generic variation of the non-zero boundary fields is
\beq
\label{eq:HolWard}
\begin{split}
\delta S \ =& \ \delta_g S + \delta_{A^I}S + \delta_{X_1}S + \delta_{b^+}S \\
=& \ \int_{\partial Y_5 = M_4}\left[ \frac{1}{2}T_{ij}\delta g^{ij} + \Jcur_I^i \delta A^I_i + \Xi \delta X_1 + \frac{1}{2}\Upsilon^{+,ij}\delta b_{ij}^+ \right] \, \vol_4 \, ,
\end{split}
\eeq
since we keep $\ma$ and $b^-$ fixed to zero in order to preserve the boundary conditions \eqref{eq:NOTwist}. As in previous uses of relations of this type, such as \cite{BenettiGenolini:2017zmu, BenettiGenolini:2018iuy}, the variation of the on-shell action is necessarily a boundary term provided that the bulk does not have any singularities or internal boundaries, or there would be additional contributions from those loci where the equations of motion are not satisfied. We focus on smooth fillings, but this could be too narrow in general. It is possible that singular gravitational fillings provide the dominant contribution to the to the saddle point approximation in \eqref{eq:AdSCFT}. If this was the case, a relevant question could then be: what kind of singularities should we allow in $Y_5$ so that the holographic Ward identity still holds? Nevertheless, note that, as in \cite{BenettiGenolini:2017zmu}, our results would not be spoiled by mild singularities, depending on the radial behaviour of the fields near them.

The boundary conditions \eqref{eq:NOTwist} relate the variation of the boundary fields $A^I$ and $X_1$ to that of the metric, since the former are clearly fixed in terms of the latter. Thus, we can write the sum of the first three contributions to \eqref{eq:HolWard} as
\begin{align}
\delta_g S + \delta_{A^I}S + \delta_{X_1}S \ = \ \frac{1}{4\kappa^2_5}\int_{M_4}\left( \mc{T}_{ij}\delta g^{ij} + \mathscr{D}_S \right) \vol_4 \, , 
\end{align}
where $\mathscr{D}_S$ is a total derivative, which we ignore assuming that $M_4$ is a closed manifold.
We can then use the expressions for the one-point functions \eqref{Tij}-\eqref{bbJ}, together with the boundary conditions, to write
\begin{align}
\mc{T}_{ij} \ =& \  4 \mexp^4_{ij} + h^1_{ij} - 4g_{ij} \left( t^{(4)} - \tfrac{1}{2} t^{(2,2)} - \tfrac{1}{8} u^{(1)}  \right)  - 6 g_{ij} \Xs^2 - 2 \mexp_{ij}^{2} t^{(2)} \nn\\
\label{eq:CurlyT}
	& + \tfrac{1}{2} \Big( \nabla^k \nabla_i \mexp^2_{jk} + \nabla^k \nabla_j \mexp^2_{ik} - \nabla^2 \mexp^2_{ij} - \nabla_i \nabla_j t^{(2)} \Big) + \tfrac{1}{2} g_{ij} \big( \mexp^2_{kl} R^{kl} \big) - \tfrac{1}{2} \mexp_{ij}^{2} R
	  \\
	& - \Big( \Xs R_{ij} + g_{ij}\nabla^2 \Xs - \nabla_i\nabla_j \Xs \Big) - \tfrac{1}{2} \Big[ \mathcal{D}^k ( \aIfour + 2 \aIfive  )_i \, {\Jb^I_{jk}} \Big] \, . \nn 
\end{align}
Here there have been explicit cancellations involving $\langle (\rd\xi^{\flat})^- , b^-_3\rangle$ so that formally this effective stress-energy tensor is the same as in \cite{BenettiGenolini:2017zmu}. Note, however, that there are implicit contributions from the antisymmetric tensor fields through $4 \mexp^4_{ij} + h^1_{ij}$ etc.

The contribution from the variation of $b^+$ is slightly subtler. We are considering variations of the metric that preserve the Riemannian structure of the manifold with the isometry $(M_4,g,\xi)$, so $\delta \xi=0$, but $\delta\xi^{\flat} \neq 0$. Moreover, $b^+ = 2\rd\xi^-$, so we also have contributions from the variation of the Hodge dual. Overall, we find
\beq
\begin{split}
\delta \left((\rd\xi^\flat)^-_{ab}\right) \ =& \ (\rd(\xi\hook \delta g))_{ab}^-  + \frac{1}{2}\sqrt{g}\epsilon^p_{\ph{p}cab}\delta g_{pq}(\rd\xi^\flat)^{qc} - \frac{1}{4}(*\rd\xi^\flat)_{ab}\, g^{mn}\delta g_{mn} \, ,
\end{split}
\eeq
and the contribution to the holographic Ward identity of the one-point function $\Upsilon^{+,ij}$ in \eqref{Ypm} evaluates to
\beq
\label{eq:bContrib}
\begin{split}
\frac{1}{2}\Upsilon^{+,ij}\delta b^+_{ij} \ =& \ \frac{1}{4\kappa^2_5}\bigg[ \left( \xi^{\flat}_{(i}(*\rd b_3^-)_{j)} - \frac{1}{2} (b_3^-)_{(i}^{\ph{(i}k}(\rd\xi^\flat)^+_{j)k}  \right)\delta g^{ij}  - \nabla_i\left( b_3^{-,ij}\delta g_{jk}\xi^k\right) \bigg] \, .
\end{split}
\eeq
As in the previous expression, we can ignore the total derivative, since we assume that $M_4$ is closed, so we can overall write
\begin{align}
\label{eq:deltaS1}
\delta S \ = \ \frac{1}{4\kappa^2_5}\int_{M_4}\widetilde{\mc{T}}_{ij}\delta g^{ij} \, .
\end{align}

We report here the values of the combinations of bosonic fields in \eqref{eq:CurlyT} from \cite{BenettiGenolini:2017zmu} with the boundary conditions \eqref{eq:NOTwist}
\begin{gather}
\mexp^2_{ij} \ = \ - \tfrac{1}{2}\left(R_{ij} - \tfrac{1}{6}Rg_{ij} \right) \, , \qquad t^{(2)} \ = \ - \tfrac{1}{6}R \, , \\
\begin{split}
- 4g_{ij} \left( t^{(4)} - \tfrac{1}{2} t^{(2,2)} - \tfrac{1}{8} u^{(1)}  \right) \ =& \  g_{ij}\Big( 8 X_2^2 + \tfrac{1}{6}RX_2 - \tfrac{1}{48}(\mc{P}+\mc{E}) - \tfrac{1}{12}\langle (\rd \xi^{\flat})^-, b^-_3 \rangle \\
&\qquad  + \tfrac{1}{8} R_{kl}R^{kl} - \tfrac{1}{36}R^2 \Big) \, .
\end{split}
\end{gather}
Substituting these expressions in \eqref{eq:CurlyT} together with the expression \eqref{eq:g4h1SUSY} and adding the contribution \eqref{eq:bContrib}, we find
\beq
\begin{split}
\widetilde{\mc{T}}_{ij} \ =& \ 3 \nabla_i\nabla_j X_2 + 2\ii \nabla_{(i}(\ma_2)_{j)} + \frac{1}{4}\nabla^2R_{ij} - \frac{1}{4}R_{ik}R^k_{\ph{k}j} - \frac{1}{4}R_{iklj}R^{kl}  \\
& + g_{ij}\bigg( -\nabla^2X_2 - \frac{1}{24}\nabla^2R- \frac{\mc{P}+\mc{E}}{48} + \frac{1}{3}\langle \rd b^-,\xi^{\flat} \rangle + \frac{1}{24}\langle (\rd \xi^{\flat})^-,b_3^- \rangle \bigg) \\
& - \frac{1}{8}\epsilon^{mnk}_{\ph{mnk}j}R_{mnli}R_k^{\ph{k}l} - \nabla_{(i}(\xi \hook b_3^-)_{j)} + \frac{1}{3}\left[2\mc{D}a^I_2 - *\mc{D}a^I_2 \right]_{(i|k|}\Jb^{Ik}_{\ph{Ik}j)} \\
& - \frac{1}{2}\mc{D}^k\left(a^I_1 + 2a^I_2\right)_{(i}\Jb^I_{j)k} -\frac{1}{8}g_{ij} \langle (\rd\xi^{\flat})^- , b^-_3\rangle  \, .
\end{split}
\eeq
Now we use the equations \eqref{eq:covDX2} and \eqref{eq:LaplacianX2} coming from supersymmetry, together with the expression for $a^I_1$ from the bosonic equations of motion, to arrive at
\beq
\begin{split}
	\widetilde{\mathcal{T}}_{ij} \ =& \ \ \ \frac{1}{4} \nabla^2 R_{ij} -\frac{1}{8}\nabla_i\nabla_j R+ \frac{1}{4}\nabla^k \nabla^l R_{jkli} - \frac{1}{4} R_{ik} R^k{}_j - \frac{1}{4} R_{iklj} R^{kl}  \\
	& - \frac{1}{6} g_{ij} \left( \mathcal{D} \aIfive \right)^{kl} \Jb^{I}_{kl} + \frac{1}{3} [ 2\mathcal{D}\aIfive - *(\mathcal{D}\aIfive) ]_{(i|k|} \Jb^{I k}{}_{j)} - ( \mathcal{D}\aIfive )_{(i|k|} \Jb^{I k}{}_{j)}  \\
	&+ \frac{1}{8}\epsilon_{j}{}^{kmn} ( 2 \nabla_k\nabla_m R_{ni} - R_{mni}{}^l R_{kl} ) \\
	&+ \frac{1}{8} g_{ij} \langle (\rd \xi^{\flat})^-, b_3^- \rangle -\frac{1}{8}g_{ij} \langle (\rd\xi^{\flat})^- , b^-_3\rangle \, .
\end{split}
\eeq
The first three lines vanish, as explained in \cite{BenettiGenolini:2017zmu}: the first line as a consequence of the contracted Bianchi identity, the second line because of the self-duality properties of $\Jb^I$, and the third line is zero after applying the Ricci identity for a rank-two covariant tensor and the first Bianchi identity. The final line, which collates the entire contribution of the antisymmetric tensors, is trivially zero. Overall, we find that
\beq
\delta S \ = \ \frac{1}{4\kappa^2_5}\int_{M_4}\widetilde{\mc{T}}_{ij}\delta g^{ij} \ = \ 0 \, .
\eeq
Let us comment on this result. At first glance, it seems that the above result holds for \emph{any} variation $\delta g^{ij}$ of the boundary metric, since we have not needed to use $\mathcal{L}_\xi \delta g=0$ as anticipated in \cite{BenettiGenolini:2017zmu}. However, we must recall that the class of boundary manifolds we are considering is restricted, as we are computing $\widetilde{\mc{T}}$ on a supersymmetric background with a fixed Killing vector field. Thus, we are considering the variation of a functional on a constrained subset of its domain, and we cannot conclude that the vanishing result holds everywhere in the domain: a generic variation of the metric would take us outside of the locus where $\widetilde{\mc{T}}$ has been evaluated.

On the other hand, we can verify that the action does indeed depend on the choice of isometry by allowing $\xi$ to vary. The vector field $\xi$ enters only in the definition of $b^+$, so it changes \eqref{eq:bContrib}, adding to it a term proportional to $\delta\xi$. The result is that \eqref{eq:deltaS1} becomes (apart from total derivative terms)
\beq
\delta S \ = \ \frac{1}{4\kappa^2_5}\int_{M_4}\left( \widetilde{\mc{T}}_{ij}\delta g^{ij} + \nabla^j(b^-_3)_{ji}\, \delta\xi^i \right) \, .
\eeq
Recall that $b^-_3$ is not determined by the boundary data, nor is its divergence. Therefore, we conclude that in absence of additional information about the structure of the bulk the on-shell action does depend on the choice of isometry, as expected from field theory.

\medskip

Note that both results have been reached in the minimal holographic renormalization scheme. However, supersymmetry may require the inclusion of additional finite counterterms, as happens with scalars \cite{Freedman:2013ryh, Bobev:2013cja, Bobev:2016nua, Freedman:2016yue, Kol:2016ucd}, or it could have even been anomalous, as pointed out and clarified in \cite{Genolini:2016sxe, Genolini:2016ecx, Papadimitriou:2017kzw, An:2017ihs, Papadimitriou:2019gel, Closset:2019ucb}.

\section*{Acknowledgments}
We would like to thank James Sparks for insightful comments on the manuscript. The work of PBG has been supported by the STFC consolidated grant ST/P000681/1. PR is funded through the STFC grant ST/L000326/1.

\bibliographystyle{./OmegaDeformationFiles/JHEP}
\bibliography{./OmegaDeformationFiles/Bib_OmegaDeformation}

\end{document}